\documentclass[aps,prb,twocolumn,amsmath,amssymb,superscriptaddress,floatfix]{revtex4-1}
\usepackage[utf8]{inputenc}
\usepackage{textcomp}
\usepackage{amsmath}
\usepackage{amssymb}
\usepackage{graphicx,graphics}
\usepackage{latexsym,verbatim}
\usepackage{dcolumn} 
\usepackage{color}
\usepackage{cancel}
\usepackage[breaklinks]{hyperref}

\usepackage{graphicx}  
\usepackage{dcolumn}   
\usepackage{bm}        

\usepackage{graphics}
\usepackage{latexsym}
\usepackage{amsmath}
\usepackage{float}
\usepackage{pstricks,pst-plot}

\begin{document}

\title{Magnetic tilting and emergent Majorana spin connection in topological superconductors}

\author{Luca Chirolli}
\affiliation{IMDEA-Nanoscience, Calle de Faraday 9, E-28049 Madrid, Spain}

\author{Francisco Guinea}

\affiliation{IMDEA-Nanoscience, Calle de Faraday 9, E-28049 Madrid, Spain}
\affiliation{School of Physics and Astronomy, University of Manchester, Manchester, M13 9PY, UK}

\begin{abstract}
Due to the charge neutral and localized nature of surface Majorana modes, detection schemes usually rely on local spectroscopy 
or interference through the Josephson effect. Here, we theoretically study the magnetic response of a two-dimensional cone of 
Majorana fermions localized at the surface of class DIII Topological Superconductors. For a field parallel to the surface the Zeeman 
term vanishes and the orbital term induces a Doppler shift of the Andreev levels resulting in a tilting of the surface Majorana cone. 
For fields larger than a critical threshold field $H^*$ the system undergoes a transition from type I to type II Dirac-Majorana cone. 
In a spherical geometry the surface curvature  leads to the emergence of the Majorana spin connection in the tilting term via an 
interplay between orbital and Zeeman, that generates a finite non-trivial coupling between negative and positive energy states. 
Majorana modes are thus expected to show a finite response to the applied field, that acquires a universal character in finite 
geometries and opens the way to detection of Majorana modes via time-dependent magnetic fields.
\end{abstract}

\maketitle

\section{Introduction}

Majorana particles are charge neutral fermions and as such they do not couple to the electromagnetic field \cite{Majorana1937}. 
In solid state systems \cite{Alicea} they are formed as zero energy equal superposition of particle and hole states, localized 
at a vortex core \cite{Ivanov2001} or at the boundary of topological superconductors \cite{Kitaev2001,Mourik1003}. Majorana 
fermions constitute a particular class of topologically protected surface states appearing at the boundary of topological states of 
matter \cite{QiRMP2011,AndoFu-review,Bernevig-book,HasanRMP2010,Chiu}, and they represent one of the basic resources in 
topological quantum computation \cite{TewariZoller2007,nayak}. Majorana states localized at the surface of two-dimensional (2D) 
and three-dimensional (3D) topological superconductors form modes that fill the entire gap \cite{Altland,SchneyderPRB2008,Kitaev}. 
In class DIII topological superconductors in $D=3$ dimensions \cite{ZhangKaneMele,Ryu,Volovik-book,Qi-PRL2009,Qi-PRB2010},  
Majorana modes form a time-reversal invariant (TRI) Dirac cone in the basis of Majorana Kramers partners. A novel class of material 
based on doped Bi$_2$Se$_3$ Topological Insulators (TI) \cite{QiRMP2011,HasanRMP2010,ZhangNatPhys2009,Fang,FuPRL2007,
ChenScience2009} has been suggested as a candidates that may realize odd-parity, TRI topological superconductivity 
\cite{FuBerg,HorPRL2010,KrienerPRL2011,peng-prb2013,SasakiPRL2011,wang,wray-np2010, venderbos2015}.  Due to 
their charge neutrality and localized character, most of the theoretical proposals and the experimental efforts to detect Majorana 
states have focused on local spectroscopy, Josephson effect, and interferometry, as a proof of their existence \cite{Beenakker,Ramon}. 
Besides, surface Majorana modes are expected to produce a strong anisotropy in the spin suscepibility \cite{ZhangPRL2009}.

In this work we study the coupling of a 2D cone of Majorana fermions localized at the surface of a class DIII topological 
superconductors to an externally applied magnetic field.  Meissner screening fixes the relevant field orientation to lay in the 
surface and the systems responds with a supercurrent. The latter gives rise to a Doppler shift of the energy levels, that results 
in a tilting of the Majorana cone along the supercurrent direction. Upon increasing the external field a threshold field, $H^*$, 
is reached for which the dispersion becomes flat along the tilting direction and the Majorana cone experiences a structural 
change from type I to type II Dirac-Majorana cone \cite{SoluyanovNature2015,BradlynScience2016}. The $H^*$ is upper 
bounded by the thermodynamic critical fields $H_c$, and for field smaller than $H^*$ the system is not expected to respond 
to the field. For highly doped small band-gap Dirac insulators hosting odd-parity SC, such as the $A_{1u}$ phase predicted 
in Bi$_2$Se$_3$ \cite{FuBerg}, $H^*$ can lie within the Meissner phase, thus making accessible the type II Dirac cone regime.  

In systems characterized by a finite surface curvature the triplet nature of the pairing leads to an additional non-trivial 
coupling of geometric origin to the external orbital field. The tilting term, involving only the momentum operator on a flat 
surface, necessarily leads to the emergence of the Majorana spin connection on curved space and adds to the Zeeman term, 
both involving the spin density operator perpendicular to the surface. The external field thus generates a finite coupling between 
occupied and empty states and a Majorana-photon coupling thus appears on curved surfaces, allowing detection of surface 
Majorana states by application of time dependent magnetic field. Our findings acquires a universal character in finite geometries 
characterized by a non-zero curvature and open the way to detection and manipulation of Majorana states via electromagnetic 
fields.

\section{The system}

We start our analysis by considering a generic example of class DIII topological superconductors, the Balian-Werthamer (BW) state 
\cite{BalianPR1963}, that describes time-reversal invariant triplet pairing of the form $({\bf p}\cdot{\bf s})is_y$. The mean field Hamiltonian 
in the Nambu basis $\psi_{\bf k}=(c_{{\bf k}\uparrow},c_{{\bf k},\downarrow},c^\dag_{-{\bf k},\downarrow},-c^\dag_{-{\bf k},\uparrow})^T$, 
with $c_{{\bf k},s}$ a fermionic state of momentum ${\bf k}$ and spin $s$, takes the form ${\cal H}=\frac{1}{2}\sum_{\bf k}\psi^\dag_{\bf k}
{\cal H}_{\rm BW}({\bf k})\psi_{\bf k}$, where the Bogoliubov de Gennes (BdG) Hamiltonian is given by
\begin{equation}\label{Eq:BWhamiltonian}
{\cal H}_{\rm BW}=\tau_z\left(\frac{{\bf p}^2}{2m}-\epsilon_F\right)+\frac{\Delta}{\hbar k_F}\tau_x{\bf p}\cdot{\bf s},
\end{equation}
with  $m$ the effective mass of the electrons, $\epsilon_F$ the Fermi energy, $k_F=\sqrt{2m\epsilon_F}/\hbar $ the Fermi momentum, 
${\bf s}$ the vector of spin Pauli matrices, $\tau_i$ Pauli matrices in the particle-hole Nambu space, and $\Delta$ the mean-field value 
of the pseudoscalar order parameter describing the BW state. The BdG Hamiltonian Eq.~(\ref{Eq:BWhamiltonian}) is time-reversal invariant, 
with ${\cal T}=is_y\hat{K}$ and $\hat{K}$ complex conjugation, and it has a full bulk gap of size $\Delta$ on the Fermi surface. 

A surface Majorana-Dirac cone is found by 
studying the subgap spectrum of the confined problem. By confining the system in the region $z>0$, a zero energy Majorana Kramers 
pair is found at $k_x=k_y=0$. Introducing the coherence length of the superconductor $\xi = \hbar v_F/\Delta$ (with $v_F=\hbar k_F/m$), 
the unnormalized wavefunction in the limit $k_F\xi\gg 1$ is given by
\begin{equation}\label{Eq:MajoWF}
\phi_s(z)=|s\rangle(|+\rangle_\tau+is|-\rangle_\tau)e^{-z/\xi}\sin(k_Fz),
\end{equation}
where $\tau_z|\pm\rangle_\tau=\pm|\pm\rangle_\tau$, $s_z|s\rangle=s|s\rangle$. For ${\bf k}\neq 0$ we project the Hamiltonian 
Eq.~(\ref{Eq:BWhamiltonian}) onto the surface states $|\phi_s\rangle$ and obtain an effective surface Hamiltonian valid  for energies 
below the gap 
\begin{equation}\label{H0}
{\cal H}^0_{\bf k}=\hbar v_\Delta(k_xs_y-k_ys_x),
\end{equation}
with $v_\Delta=\Delta/\hbar k_F$, that represents a massless Dirac-Majorana Hamiltonian describing the surface physics.

\section{Coupling to a magnetic field}

The magnetic field is introduced in Eq.~(\ref{Eq:BWhamiltonian})  by minimal coupling substitution, ${\bf p}\to {\bf p}+e{\bf A}\tau_z$ 
in the kinetic energy, and the gap acquires a dependence on the position. For weak magnetic field in the Meissner 
regime the vector potential and the gap change over distances on order of the penetration depth $\lambda$ and the electromagnetic 
properties of the pseudoscalar order parameter $\Delta$ are similar to those of a scalar $s$-wave order parameter 
\cite{Abrikosov,BalianPR1963,Sigrist}. We then assume a constant hard gap that is not affected by the magnetic field. 
Neglecting the diamagnetic ${\bf A}^2$ term the coupling to the magnetic field reads
\begin{equation}\label{HpA}
{\cal H}_{B}=\frac{e}{2m}({\bf p}\cdot{\bf A}+{\bf A}\cdot{\bf p})+\frac{1}{2}g\mu_B{\bf s}\cdot{\bf B}, 
\end{equation}
both for particles and holes, where the second term describes Zeeman coupling to the total induction field ${\bf B}=\nabla\times {\bf A}$, 
with $g$ the electron $g$-factor and $\mu_B$ the Bohr magneton. The BdG Hamiltonian transforms under Particle-Hole (PH) symmetry 
as $U_{\cal C}^\dag {\cal H}^*(-{\bf k},H) U_{\cal C}= -{\cal H}({\bf k},H)$, with $U_{\cal C}=s_y\tau_y$. Consequently, the spectrum has 
the following symmetry $\epsilon^n({\bf k},H)=-\epsilon^n(-{\bf k},H)$. 

\begin{figure}[t]
\begin{center}
\includegraphics[width=0.47\textwidth]{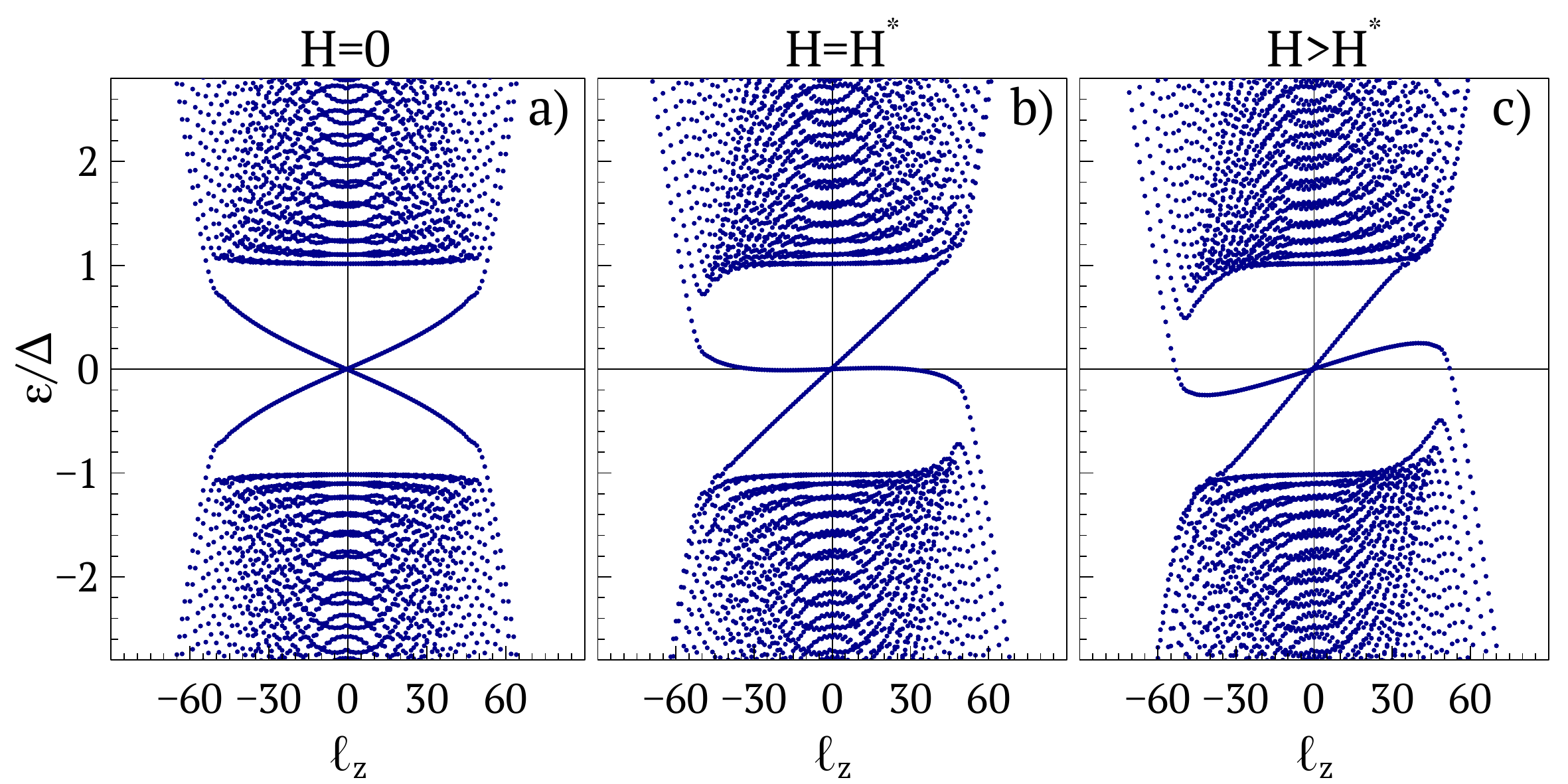}
\caption{Evolution with the field of the spectrum on a cylindrical geometry for the case $k_z=0$ versus angular momentum 
$\ell_z=j_z-1/2$. The applied field is $H=0$ in a), $H=H^*$ in b) and $H>H^*$ in c) and it exponentially decays towards the center 
of the cylinder on a scale $\lambda=0.06~R$. PH symmetry requires that $\epsilon^n_{\ell_z}(H)=-\epsilon^n_{-\ell_z-1}(H)$.
\label{Fig1}}
\end{center}
\end{figure}

When an external field $H$ is applied to the system Meissner screening forces the external field to lay in the plane defined by the 
surface of the system. The Zeeman term involves then only in-plane spin components, whose matrix elements on the states 
Eq.~(\ref{Eq:MajoWF}) are zero. In the Landau gauge the vector potential corresponding to the bulk screened field reads 
${\bf A}=(0,H\lambda e^{-z/\lambda},0)$, with the field ${\bf B}$ pointing about the $x$-direction. Having no left structure in spin and 
particle-hole space, the projection onto the Majorana cone of ${\cal H}_{B}$ can only amount to a diagonal momentum-dependent term,
\begin{equation}\label{HsurfA}
{\cal H}_{\bf k}=\hbar v_\Delta(k_xs_y-k_ys_x)+\hbar v_Hk_y,
\end{equation}
where the field-induced velocity is given by $v_H=e \langle A_y\rangle/m\simeq e A_y(0)/m =eH\lambda/m$, in the limit $k_F\xi\gg 1$, 
and  $\langle\ldots\rangle$ stands for the expectation value on the Majorana wavefunction Eq.~(\ref{Eq:MajoWF}). This term describes 
a Doppler shift of the energy of the Majorana modes due to the superfluid velocity generated by the field  and accounts for 
the Galilean transformation of the energy levels from the comoving to the lab frame \cite{deGennes}. The result is a tilting of the bands 
described by the energy dispersion $\epsilon_{{\bf k},\pm}=\hbar v_Hk_y\pm \hbar v_\Delta k$. Upon increasing the field the tilting 
becomes more and more pronounced, until a threshold field is reached, where one of the two energy branches becomes flat at the 
Landau critical velocity 
$v_H=v_\Delta$,
\begin{equation}\label{Eq:ThrField}
H^*=\eta\frac{\Phi_0}{\pi\xi\lambda}=\eta H_c.
\end{equation}
In the second equality we express the threshold field in terms of the thermodynamic critical field $H_c=\Phi_0/(2\pi \xi\lambda)$  
\cite{deGennes}. For type II superconductor $H_c$ lays between the lower and upper critical fields, $H_{c1}< H_c<H_{c2}$, 
with $H_{c1}=\Phi_0/(4\pi\lambda^2)\log(\lambda/\xi)$ and $H_{c2}=\Phi_0/(2\pi \xi^2)$, respectively \cite{deGennes}.  At $H=H^*$ 
the Majorana cone becomes flat along the line $k_x=0$ and a transition from type I to an overtilted type II Dirac-Majorana cone takes 
place, that is characterized by a formally diverging number of states at zero energy. In Fig.~\ref{Fig1} we plot the band structure of 
the Hamiltonian Eq.~(\ref{HamTI}) in a cylindrical geometry with a screened field. From left to right the field is increased from $H=0$ 
to $H>H^*$ and the tilting of the bands is evident, thus confirming the prediction of the low energy model. The parameter $\eta$ in 
Eq.~(\ref{Eq:ThrField}) is the ratio of the actual Majorana velocity $v_\Delta$ and $\Delta/\hbar k_F$, and it captures the discrepancy 
between the actual Hamiltonian and the BW state Eq.~(\ref{Eq:BWhamiltonian}) ($\eta=1$ for the BW state).

\subsection{Odd-parity superconductor}

A relevant example is represented by odd-parity superconductors realized in doped TI, like Bi$_2$Se$_3$, or small band-gap doped 
Dirac insulator. These systems are described by the BdG Hamiltonian \cite{FuBerg} 
\begin{equation}\label{HamTI}
{\cal H}=\tau_z(E_G\sigma_x+v\sigma_z(k_xs_y-k_ys_x)+v_zk_z\sigma_y)+\Delta \tau_x\sigma_ys_z,
\end{equation}
where $\sigma_i$ are Pauli matrices spanning the subspace of two orbitals with $p_z$-like symmetry, $E_G$ is the 
insulating band-gap, and $v$ the Dirac velocity. The BW Hamiltonian Eq.~(\ref{Eq:BWhamiltonian}) can be seen as the 
low energy limit of the Hamiltonian Eq.~(\ref{HamTI}) and can be obtained upon projection on the conduction band. 
Introducing euclidean Dirac matrices, 
$\gamma^\mu=(\gamma^0,\gamma^1,\gamma^2,\gamma^3)=(\sigma_x,-\sigma_ys_y,\sigma_ys_x,\sigma_z)$ and 
$\gamma^5=\sigma_ys_z$, and assuming isotropic velocity $v$, minimal coupling substitution ${\bf p}\to{\bf p}+\tau_z{\bf A}$ 
produces the coupling term ${\cal H}_{\bf A}=-iev\gamma^0\gamma^iA_i$. Defining ${\cal H}_0=\tau_z(E_G\gamma^0-\mu)$, 
${\cal H}_{\bf p}=-iv\gamma^0\gamma^ip_i\tau_z$, and ${\cal H}_\Delta=\Delta\tau_x\gamma^5$, the projected coupling is given by 
\begin{equation}
\tilde{\cal H}={\cal H}_{\rm BW}+{\cal H}_{{\bf p}\cdot{\bf A}}+{\cal H}_Z, 
\end{equation}
where ${\cal H}_{\rm BW}=\langle\psi_c|{\cal H}_0-{\cal H}_{\bf p}{\cal H}_{0}^{-1}{\cal H}_{\bf p}
-{\cal H}_{\Delta}{\cal H}_{0}^{-1}{\cal H}_{\bf p}|\psi_c\rangle$ is the BW 
state Hamiltonian Eq.~(\ref{Eq:BWhamiltonian}) with $|\psi_c\rangle\langle\psi_c|$ conduction band projector, $m=(\mu+E_G)/2v^2$ 
and $k_F=(\mu+E_G)/2v$, ${\cal H}_{{\bf p}\cdot{\bf A}}$ is given by the first term in Eq.~(\ref{HpA}) in the a gauge for which 
$\nabla\cdot{\bf A}=0$, and ${\cal H}_Z=g^*\mu_Bs_iB_i$ is the Zeeman term with the effective $g$-factor 
$g^*=m_{\rm e}/m+g_{{\rm Bi}_2{\rm Se}_3}/2$, $m_{\rm e}$ the bare electron mass, and we have included the material $g$-factor. 
Importantly, no coupling between the gap matrix and the vector potential arise at second order, 
$\langle\psi_c|{\cal H}_{\bf A}{\cal H}_0^{-1}{\cal H}_\Delta|\psi_c\rangle=0$. The velocity of the Majorana modes derived 
from Eq.~(\ref{HamTI}) is $v_\Delta\simeq vE_G\Delta/\mu^2$ \cite{HsiehPRL2012}. Upon identifying $m=\mu/2v^2$, one has 
$\eta=E_G/\mu$. Thus, by increasing the chemical potential $\mu$ it is possible to shift $H^*$ in the Meissner regime.

In a upcoming work \cite{ChirolliArxiv180209204} we will show that for fields larger than the threshold field, $H>H^*$, the surface 
Majorana modes carry a finite current that contributes to Meissner screening and gives rise to an additional surface diamagnetic 
moment, that can be detected in  orbital magnetic susceptibility measurements. The analysis so far presented fully makes sense 
for weak $H^*$ ($\eta\ll 1$) in the Meissner phase, where the bulk completely expels the external field. For $\eta\sim 1$ vortices 
start to enter the system before $H^*$ is actually reached. In this case, the field can be assumed constant, $A_y=-Hz$, and the 
tilting field shifts to $H_{c2}=\Phi_0/(\pi\xi^2)$, for which superconductivity is completely destroyed. Besides, the application of high 
fields may make favourable the condensation of other possible multicomponent superconducting order parameters \cite{Chirolli-2017}, 
that due to the unconventional character of the pairing are expect to affect the magnetic phase diagram, and the entire analysis 
ceases to be valid.

\section{Curvature effects}
 
We now study the modification of the Majorana cone in curved, finite size geometries. For simplicity, we will set $\eta=1$ in the following. 
In cylindrical and spherical geometry it is possible to choose the self-consistent vector potential to have only non-zero azimuthal 
component ${\bf A}=\hat{\boldsymbol{\phi}}A_\phi(r)$, with $\nabla\cdot{\bf A}=0$. The coupling to the magnetic field can be written as
\begin{equation}\label{Hpa-sphere}
{\cal H}_{B}=\frac{e}{m}A_\phi({\bf r})p_\phi+g^*\mu_B{\bf B}({\bf r})\cdot{\bf s}.
\end{equation}
Both in cylindrical and spherical geometries the $z$-component of the total angular momentum $\hat{J}_z=\hat{L}_z+\hbar s_z/2$, with 
$\hat{L}_z=-i\hbar \partial_\phi$ the $z$-component of the orbital angular momentum $\hat{\bf L}$, is a constant of motion and the 
spectrum can be labeled by $j_z$.  On the cylinder we find that the surface Hamiltonian is given by the counterpart of 
Eq.~(\ref{HsurfA}) for the surface of the cylinder (results not shown).

\begin{figure}[t]
\includegraphics[width=0.45\textwidth]{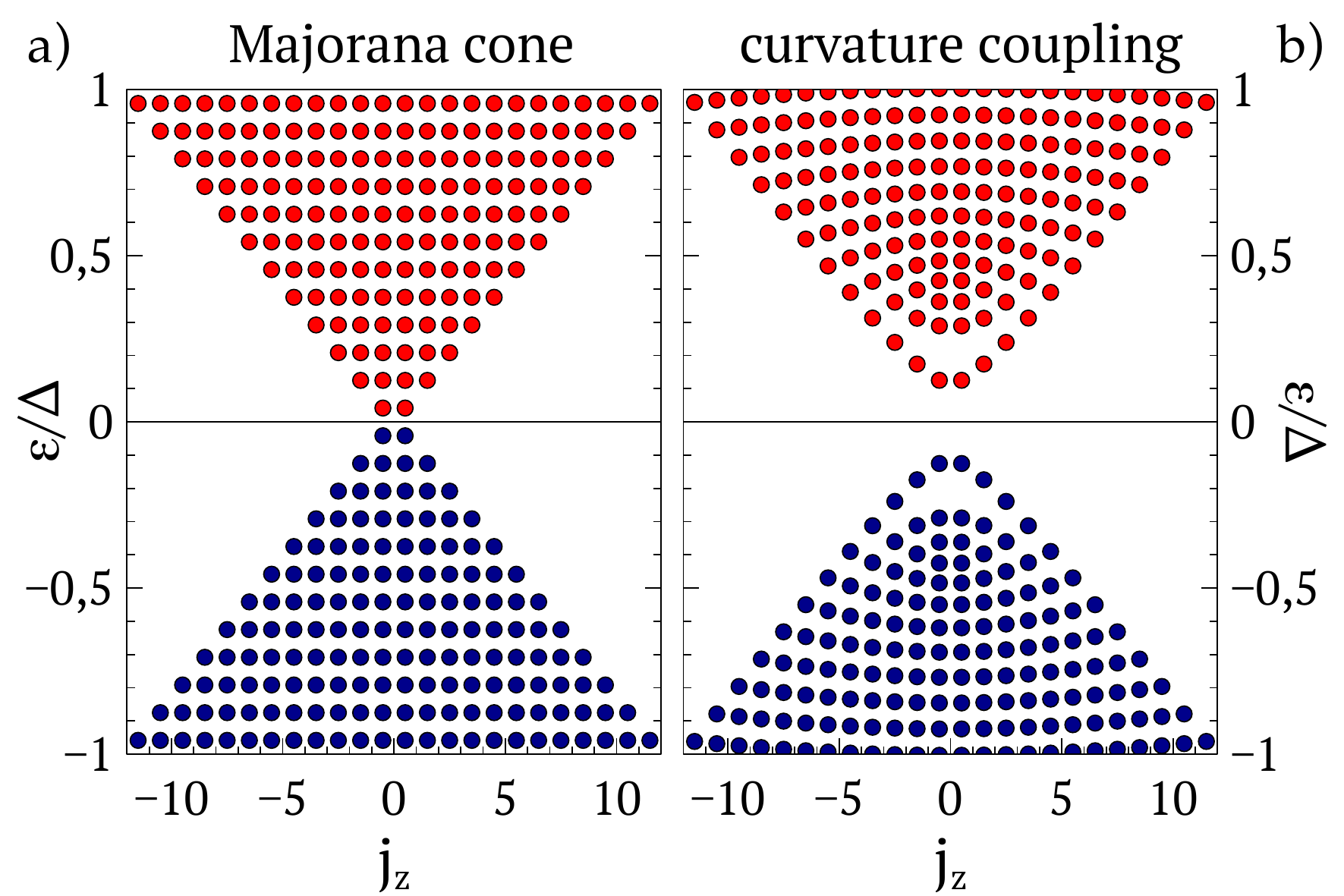}
\caption{Sketch of the surface Andreev states of the BW states in a spherical geometry. a) Surface Majorana cone without magnetic field, 
b) effect of the matrix elements Eq.~(\ref{CouplingSz}) on the spectrum. Large values of the external field have been chosen to highlight the effect.
\label{Fig:DiracSphere}}
\end{figure}

On a sphere of radius $R$, ${\cal H}$ commutes with $\hat{\bf J}^2$, $\hat{J}_z$, and ${\cal C}=\tau_z({\bf s}\cdot\hat{\bf L}+\hbar)$, 
with eigenvalues $j(j+1)\hbar^2$, $j_z\hbar$, and $\kappa=\pm(j+1/2)\hbar$. Thanks to $\{{\cal C},\Theta\}=0$, with $\Theta=\tau_y{\cal T}$ 
the particle-hole symmetry operator, the sign of $\kappa$ can be chosen to label  the sign of the eigenenergies. In turn, $[{\cal C},s_z]\neq 0$ 
implies that $s_z$ can generate finite matrix elements between the positive and negative energy states. Writing 
$p_\phi=\hbar(j_z-s_z/2)/(r\sin\theta)$ and $A_\phi({\bf r})=A(r)\sin\theta$, for large $j_z$ the spectrum is given by 
\begin{equation}
\epsilon_{j,j_z}^\pm=\pm\frac{\Delta}{k_FR}(j+1/2)+j_z\frac{\hbar e\lambda H}{mR}\qquad {\rm for}~|j_z|\gg 1/2. 
\end{equation}
In absence of the external field every state 
labeled by $j$ is $2j+1$ degenerate, corresponding to the values of $j_z$, with $|j_z|\leq j$ (see Fig.~\ref{Fig:DiracSphere}a)). As for the 
planar and cylindrical geometry the field produces a tilting of the spectrum and a threshold field $H^*$ arises, whose value agres well with 
Eq.~(\ref{Eq:ThrField}) (see Appendix Sec. \ref{App:CouplingToField}). For small $j_z$ the role of the spin becomes important and both the 
orbital and Zeeman term act as relevant perturbations to the Majorana modes. The surface contribution of the Zeeman term reads 
\begin{equation}
{\cal H}_Z=g^*\mu_B\frac{3H}{2}\left[\sin\theta~\hat{\boldsymbol{\theta}}-\frac{\lambda}{2R}
\cos\theta ~\hat{\bf r}\right]\cdot{\bf s}
\end{equation}
The polar term acts in-plane and its matrix elements are negligible (see Appendix Sec.~\ref{App:Zeeman}), so that the relevant 
perturbation provided by the magnetic field reads  
\begin{equation}
{\cal H}_B=-\mu_B\frac{A(r)}{r}\left(s_z-2g^*\cos\theta~\hat{\bf r}\cdot{\bf s}\right).
\end{equation}
This term is even under parity so that it generates finite matrix elements between states that differ by an odd number of angular momentum 
quanta. In particular, it generates matrix elements between states of positive energy and quantum numbers $(j,j_z)$ and states of negative 
energies and quantum numbers $(j\pm 1,j_z)$. A partial cancelation takes place between the orbital term and part of the Zeeman term and 
the net matrix elements read (see Appendix Sec.~\ref{App:OrbitalZeeman}) 
\begin{equation}\label{CouplingSz}
\langle\Psi^+_{j,j_z}|{\cal H}_B|\Psi_{j\pm 1,j_z}^-\rangle=\frac{\tilde{g}\mu_B A(R)}{R}
\frac{\sqrt{(j+1/2\pm 1/2)^2-j_z^2}}{j+1/2\pm 1/2},
\end{equation}
with $A(R)=3H\lambda/2$ on the sphere surface and $\tilde{g}=g^*-1/2$. 
Analogously, the Zeeman term, whose relevant component is $s_z$, has similar finite matrix elements and couples to the Majorana cone. 
It becomes clear that the orbital magnetic field adds to the Zeeman term and acts as a relevant perturbation that opens a gap in the Majorana 
cone. The resulting spectrum without the tilting term is shown Fig.~\ref{Fig:DiracSphere}b). The magnetic field increasing the splitting between 
positive and negative energy states, as a result of a non-trivial coupling.

\subsection{Majorana Spin Connection}

The matrix elements Eq.~(\ref{CouplingSz}) hide a profound geometric origin. The Majorana modes in absence 
of the field satisfy a 2D Dirac equation, that on a planar metric is given by Eq.~(\ref{H0}). Introducing Dirac matrices 
$\gamma^i\equiv(\gamma^0,\gamma^1,\gamma^2)=(-is_z,s_y,-s_x)$, the two-component Majorana spinor 
$\psi$ satisfies the Dirac equation on a flat Minkowski space $\gamma^i\partial_i\psi=0$. Parallel transport on the surface of a sphere 
dictates the way the derivative $\partial_i$ becomes replaced by the covariant derivative $D_\mu=\partial_\mu+\Gamma_\mu$. The 
spin connection, related to the intrinsic curvature of the metric, necessarily emerges at the boundaries of a TSC, and the Majorana spinor 
satisfies the Dirac equation on the curved space-time $S^2 \times R$ \cite{Nakahara,GonzalezPRL1992,ParentePRB2011},
\begin{equation}
\gamma^\mu(\partial_\mu+\Gamma_\mu)\psi=0,
\end{equation}
where $\gamma^\mu=\gamma^ie_i^{~\mu}$ are rotated Dirac matrices satisfying $\{\gamma^\mu,\gamma^\nu\}=2g^{\mu\nu}$, 
where the curved metric $g_{\mu\nu}=\eta_{ij}e^i_{~\mu}e^j_{~\nu}$ replaces the Minkowski metric $\eta_{ij}$ via the tetrads  
$e^i_{~\mu}\equiv\partial x^i/\partial x^\mu$. The spin connection $\Gamma_\mu$ is given by $\frac{i}{2}\Gamma^{i~~j}_{~\mu}\Sigma_{ij}$, 
where $\Sigma_{ij}=\frac{i}{2}[\gamma_i,\gamma_ j]$ are the generators of the spinorial representation of the Lorentz group, and the connection 
coefficients are given by $\Gamma^{i~~j}_{~\mu}=e^i_{~\nu}\nabla_\mu e^{j\nu}$. On the surface of the sphere the metric takes the form 
$g_{\mu\nu}={\rm diag}(-1,R^2,R^2\sin^2\theta)$ and by reading off the tetrads the spin connection is found to be 
\begin{equation}
\Gamma_\mu=-\frac{i}{2}s_3\cos\theta ~\delta_{2\mu}.
\end{equation}
The resulting Dirac Hamiltonian on the surface of the sphere takes the form \cite{AbrikosovJr},
\begin{equation}\label{HdiracS}
{\cal H}^0_{\rm surf}=\frac{\Delta}{k_FR}\left(-is_2\left(\partial_\theta+\frac{\cot\theta}{2}\right)+i\frac{s_1}{\sin\theta}\partial_\phi\right),
\end{equation}
and it gives rise to a spectrum analogous to that shown in Fig.~\ref{Fig:DiracSphere}a), whose exact analytical form can be found 
in Ref.~\cite{AbrikosovJr} (see also the Appendix \ref{App:DiracSurface}). Due to the presence of the spin connection the Dirac 
equation on a curved surface has no zero eigenvalues. 
 
It is clear that finite matrix elements between positive and negative energy states of the Hamiltonian Eq.~(\ref{HdiracS}) can only arise 
via the third Pauli matrix $s_3$ and that it is reasonable to conclude that in a curved geometry the vector potential has to couple to the 
spin connection. Alternatively, one may wonder how the Dirac Hamiltonian Eq.~(\ref{HsurfA}), describing the coupling of a Majorana spinor 
to a magnetic field that lays on the surface, changes when the surface acquires a finite curvature. On a flat Minkowski space with a in-plane 
magnetic field ${\bf B}=\nabla\times{\bf A}$ pointing about the $x$ direction we can write Eq.~(\ref{HsurfA}) as
\begin{equation}\label{Hdirac2D-A}
\left(\gamma^i+\gamma^0a^i\right)\partial_i\psi=0,
\end{equation}
where $a^i=\frac{e}{m v_\Delta}\langle A^y\rangle \delta^i_{y}$ and the only non-zero component of $A^i$ is fixed to lay in the plane, 
orthogonally to the magnetic field. The Majorana equation on the surface of the sphere is then obtained by introducing the covariant derivative
\begin{equation}\label{Hdirac2D-A-S}
\left(\gamma^\mu+\gamma^0a^\mu\right)(\partial_\mu+\Gamma_\mu)\psi=0,
\end{equation}
where, analogously to $\gamma^\mu$, we have $a^\mu=a^ie_i^{~\mu}$. Writing $a^\phi=\frac{e}{m}A_\phi(R)/(R \sin\theta)$ with 
$A_\phi(R)=A(R)\sin\theta$, it follows that the Dirac Hamiltonian acquires a non-trivial term due 
to the coupling to the magnetic field
\begin{equation}
{\cal H}_{\rm surf}={\cal H}^0_{\rm surf}+\frac{\Delta}{k_FR}\frac{H}{H^*}\left(-i\partial_\phi-\frac{s_3}{2}\cos\theta\right).
\end{equation}
 we see that the spin connection introduces the relevant perturbation $s_3\cos\theta$. 
Introducing the half-integer total angular momentum eigenvalues $l=1/2,3/2,\ldots$, with projection along the polar axis 
$m=-l,\ldots,l$, the eigenvalues of the Dirac operator Eq.~(\ref{HdiracS}) read $\epsilon=\pm(l+1/2)$. Eigenfunctions 
can be written in terms of Jacobi polynomials and can be labeled by $l$, $m$, and the $\sigma={\rm sign}(\lambda)$, 
$|\Upsilon^\sigma_{lm}\rangle$. The spin connection then gives rise to the following matrix elements (see Appendix 
\ref{App:DiracSurface})
\begin{eqnarray}
\langle\Upsilon^+_{lm}|s_3\cos\theta|\Upsilon^-_{l'm'}\rangle&=&i\delta_{mm'}\frac{\sqrt{(l+1)^2-m^2}}{2(l+1)}
\delta_{l',l+1}\nonumber\\
&-&i\delta_{mm'}\frac{\sqrt{l^2-m^2}}{2l}\delta_{l',l-1},~~~
\end{eqnarray}
Upon identifying $(l,m)$ with $(j,j_z)$ we clearly see that the Majorana 
spin connection reproduces the matrix elements Eq.~(\ref{CouplingSz}), that generate transitions between empty and occupied 
states differing by one unit of angular momentum, $l'=l\pm 1$,  and give rise to a finite response to the applied field. 

\section{Conclusions}
We studied the coupling of a cone of Majorana modes localized at the surface of a class DIII topological superconductor to 
an externally applied magnetic field. Upon taking account of Meissner screening and the associated supercurrent flow we found 
a Doppler-shift mediated tilting of the Majorana cone that mirrors the entire spectrum modification. On a spherical geometry 
the surface curvature allows for a relevant coupling that is mediated by both the orbital and Zeeman terms and leads the emergence of the Majorana spin connection. 
The effect acquires a universal character in finite size systems that are topologically equivalent to the sphere. Our 
findings open the way to detection and manipulation of Majorana modes with time-dependent magnetic fields.

\section{Acknowledgments} 

The authors acknowledge very useful discussions with B. Bradlyn, F. de Juan, I. Grigorieva, and A. K. Geim. 
L.C. and F.G. acknowledge funding from from the European Union's Seventh Framework Program (FP7/2007-2013) 
through the ERC Advanced Grant NOVGRAPHENE (GA No. 290846), L.C. acknowledges funding from the Comunidad de 
Madrid through the grant MAD2D-CM, S2013/MIT-3007. F.G. acknowledges funding from the European Commission under 
the Graphene Falgship, contract CNECTICT-604391.

\appendix

\section{Geometry dependent screening}
\label{App:Meissner-Geometry}

\begin{figure}[t]
\begin{center}
\includegraphics[width=0.35\textwidth]{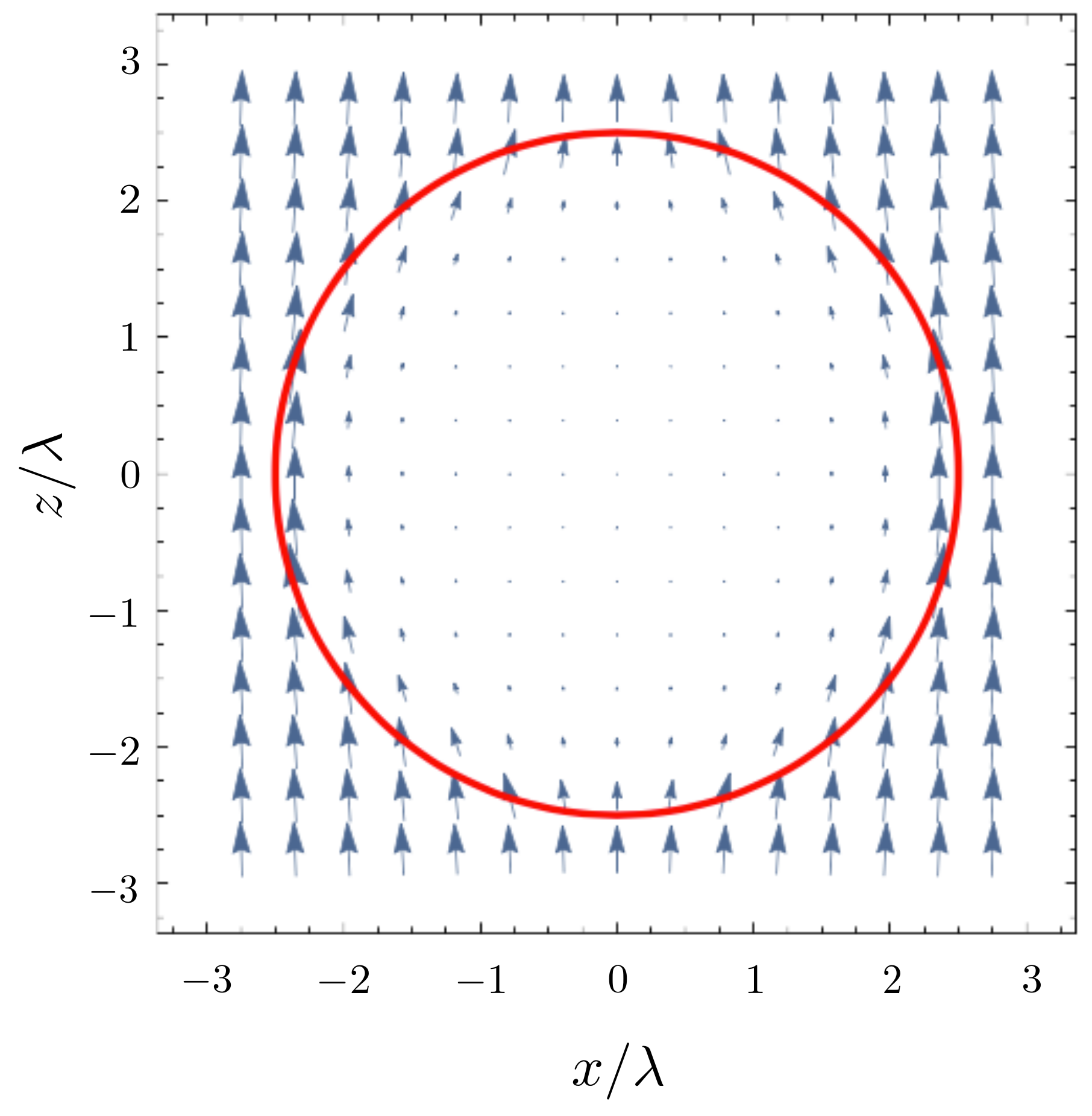}
\caption{Screened full induction field ${\bf B}$ associated to the Meissner effect for $R=10 \lambda$.
\label{Fig:SM}}
\end{center}
\end{figure}

We consider the screening of an external field $H$ by a type II superconductors of spherical geometries. 
In general we have to solve a boundary problem, inside the boundary the vector potential ${\bf A}$ has a 
mass term $1/\lambda^2$ and outside the boundary the mass term is zero. The vector potential can always 
be chosen to satisfy $\nabla\cdot{\bf A}=0$. We can choose the vector potential to be purely azimuthal, 
${\bf A}=\hat{\boldsymbol{\phi}}A_\phi(r)$. We then have to solve the following equation
\begin{equation}\label{App:Meissner-Eq}
\nabla\times\nabla\times{\bf A}=\left\{
\begin{array}{cc}
-\frac{1}{\lambda^2}{\bf A} & r<R, \vspace*{0.2cm} \\
0 &  r>R,
\end{array}
\right.
\end{equation}
with the requirement that the vector potential at sufficiently long distance from the sphere behaves as
\begin{equation}
\lim_{r\gg R}(A_r,A_\theta,A_\phi)\to\left(0,0,\frac{Hr}{2}\sin\theta\right).
\end{equation}
We write $A_\phi=f(r)\sin\theta$ 
for $r\leq R$, and $A_\phi=g(r)\sin\theta$ for $r> R$. Continuity of the magnetic induction field 
${\bf h}=\nabla\times{\bf A}$ at $r=R$ requires that $f(R)=g(R)$ and $f'(R)=g'(R)$. 
For $r<R$ the function $f(r)$ has to satisfy
\begin{equation}
\frac{f}{\lambda^2}=\frac{\partial^2f}{\partial r^2}+\frac{2}{r}\frac{\partial f}{\partial r}-\frac{2f}{r^2},
\end{equation}
whose solution reads $Af(r)$, with
\begin{equation}
f(r)={\rm Im}\left[y_{-2}(ir/\lambda)\right]
\end{equation}
and $y_\nu(z)$ a spherical Bessel function of the second kind. For $r>R$ the solution is
\begin{equation}
g(r)=\frac{H r}{2}+\frac{C}{r^2},
\end{equation}
with $A$ and $C$ real coefficients. By matching the solutions at $r=R$ we find
\begin{eqnarray}
A&=&H\frac{3R}{2(2f(R)+f'(R)R)},\\
C&=&H\frac{R^3(f(R)-Rf'(R))}{2(2f(R)+f'(R)R)}.\\ \nonumber
\end{eqnarray}
The full induction field ${\bf B}=\nabla\times{\bf A}$ then reads
\begin{equation}
{\bf B}=\left\{
\begin{array}{cc}
A\cos\theta\frac{2f(r)}{r}\hat{\bf r}-A\sin\theta\left[\frac{f(r)}{r}+f'(r)\right]\hat{\boldsymbol{\theta}} & r<R, \vspace*{0.2cm} \\
\cos\theta\frac{2g(r)}{r}\hat{\bf r}-\sin\theta\left[\frac{g(r)}{r}+g'(r)\right]\hat{\boldsymbol{\theta}} &  r>R,
\end{array}
\right.
\end{equation}
where $\hat{\bf r}$ and $\hat{\boldsymbol{\theta}}$ are unit vectors along the radial and polar directions, respectively.  
Approximating the $-{\rm Im}\left[y_{-2}(iz)\right]=e^z/(2z)$ it follows that the field for $r-R\to 0^-$ reads
\begin{equation}
{\bf B}\simeq \frac{3H}{2}\left[\sin\theta~\hat{\boldsymbol{\theta}}-\frac{\lambda}{2R}\cos\theta ~\hat{\bf r}\right].
\end{equation}
The full induction field is plotted in Fig.~\ref{Fig:SM}. We see that the field at the equator is parallel to the surface of the 
sphere and it has full strength $\sim H$, whereas at the poles it is orthogonal to the surface and it has its strength reduced 
by a factor $\lambda/R$ due to Meissner screening.

\section{BW state on the sphere}
\label{Sec:BWsphere}

We now study a purely finite size system on a sphere geometry. The BdG Hamiltonian is nothing more than the relativistic Dirac 
equation with a {\it mass} term that depends on momentum, $p^2/2m-\epsilon_F$, and the resulting second order differential 
equation for a spherical boundary can be easily solved by recalling the relativistic central force problem. The subgap spectrum 
is found by imposing the wavefunction to be zero at the boundary, $\psi(R)=0$. 

The problem can be reduced to a set of two coupled equations for the radial part of the wavefunction. It is possible to form a set of 
observables that commute with the Hamiltonian, $\hat{\bf J}^2$, $\hat{J}_z$ with quantum numbers $j=1/2,3/2,\ldots$ and $|j_z|\leq j$, 
and the operator $\hat{K}=\tau_z({\bf s}\cdot\hat{\bf L}+\hbar)$, with eigenvalue $-\hbar\kappa$, where $\hat{\bf L}$ is the orbital 
angular momentum operator. The operator $\hat{K}$  plays an important role, in that it divides the spectrum in positive and negative 
energy eigenstates, according to the sign of $\kappa=\pm(j+1/2)$. Thus, we denote the eigenstates of the Hamiltonian as 
$|\Psi^\alpha_{j,j_z}\rangle$, with $\alpha={\rm sign}(\kappa)$.

Separately, the electron and hole wavefunction are eigenstates 
of $\hat{\bf L}^2$, with quantum numbers $l_e$ and $l_h$, respectively.
The wavefunction is written as
\begin{equation}
\Psi=\left(\begin{array}{c}
f(r){\cal Y}^{j_z}_{j,l_e}(\theta,\phi),\\
g(r){\cal Y}^{j_z}_{j,l_h}(\theta,\phi),
\end{array}\right)
\end{equation}
with ${\cal Y}^{j_z}_{j,l}(\theta,\phi),$ normalized spin-angular function. For a given eigenvalue $j=1/2,3/2,\ldots$, the operator $K$ 
has eigenvalues $-\hbar \kappa$, with $\kappa=\pm(j+1/2)$. The spin-angular function are eigenstates of $\hat{\bf L}$ with eigenvalue 
$l$, and it follows that for $\kappa=j+1/2$ we have that $l_e=j+1/2$ and $l_h=j-1/2$, whereas for $\kappa=-j-1/2$ we have that $l_e=j-1/2$ 
and $l_h=j+1/2$. The relevant spin-angular wavefunction describing electron and hole components then read
\begin{equation}
{\cal Y}^{j_z}_{j,j\mp 1/2}=\left(\begin{array}{c}
\pm \sqrt{\frac{j+1/2\pm j_z\mp 1/2}{2j+1\mp 1}}Y^{j_z-1/2}_{j\mp 1/2}\\
\sqrt{\frac{j+1/2\mp j_z\mp 1/2}{2j+1\mp 1}}Y^{j_z+1/2}_{j\mp 1/2}\\
\end{array}
\right).
\end{equation}
The problem is thus reduced to a coupled set of equations for the $f$ and $g$ functions. 
Defining $L^\pm_\kappa=(\partial_r\mp \kappa/r)$ we can cast the problem in the form
\begin{equation}
\left(\begin{array}{cc}
-\frac{\hbar^2}{2m}D_\kappa-\mu-E & i\hbar v_\Delta L^+_{\kappa-1}\\
i\hbar v_\Delta L^-_{\kappa+1} & \mu+\frac{\hbar^2}{2m}D_{\kappa-1}-E
\end{array}\right)\left(\begin{array}{c}
f\\g\end{array}\right)=0
\end{equation}
where we have defined $D_\kappa=\partial^2_r+(2/r)\partial_r-\kappa(\kappa+1)/r^2$, that satisfies 
$D_\kappa\equiv L^+_{\kappa-1}L^-_{\kappa+1}$ and $D_{\kappa-1}=L^-_{\kappa+1}L^+_{\kappa-1}$. 
Introducing the spherical Bessel functions of the first kind $j_n(x)=\sqrt{\frac{\pi}{2x}}J_{n+1/2}(x)$, with 
$J_{n+1/2}$ is a Bessel function of the first kind and $n$ integer, and recalling the recurrence properties 
of the spherical Bessel functions 
\begin{eqnarray}\label{Eq:RecBesselSphere}
L^-_{n+1}j_n(qr)&=&q j_{n-1}(qr),\\
L^+_{n}j_n(qr)&=&-q j_{n+1}(qr),
\end{eqnarray}
the eigenfunctions are given by spherical Bessel functions $f=Aj_{\kappa}(kr)$ and $g=Bj_{\kappa-1}(kr)$. 
We can compactly write the wavefunction as
\begin{equation}
\Psi_{j,j_z}^{\alpha}=\left(\begin{array}{c}
Af_{l^\alpha_e}(r){\cal Y}^{j_z}_{j,l^\alpha_e}(\theta,\phi)\\
Bg_{l^\alpha_h}(r){\cal Y}^{j_z}_{j,l^\alpha_h}(\theta,\phi)
\end{array}\right).
\end{equation}
The action of the operator $\hat{K}$ on these states is 
$\hat{K}\left|\Psi_{j,j_z}^{\alpha}\right\rangle=-\alpha(j+1/2)\left|\Psi_{j,j_z}^{\alpha}\right\rangle$.

For the positive energy eigenstates we have $\kappa=j+1/2\equiv \ell$, $l_e=\ell$ and $l_h=\ell-1$, 
so that we can write $f(r)=Aj_{\ell}(kr)$ and $g(r)=Bj_{\ell-1}(kr)$, in terms of spherical Bessel functions. 
The wavevector $k$ is given by the solution of the determinantal equation
\begin{equation}
\epsilon^2-\left(\epsilon_F-\frac{\hbar^2k^2}{2m}\right)^2-\hbar^2v_\Delta^2k^2=0,
\end{equation}
with $A=iv_\Delta kB(\epsilon+\epsilon_F-\hbar^2k^2/(2m))^{-1}$. For a given energy $\epsilon$ within 
the superconducting gap, $|\epsilon|<\Delta$, the possible values of $k$ for which the wavefunction 
vanishes at the origin $r=0$ are given by
\begin{equation}
k_\pm=\pm k_F+i\frac{\sqrt{2}}{\xi}\sqrt{1-\frac{\epsilon^2}{2\Delta^2}}.
\end{equation}
The decaying length of these states in the bulk is the superconducting coherence length $\xi=\hbar v_F/\Delta$, 
with $v_F=\hbar k_F/m$, satisfying $k_F\xi=2\epsilon_F/\Delta\gg 1$. By imposing the boundary condition $\psi(R)=0$ 
we find the spectrum of the problem. For large $k_FR$  the surface states spectrum below the gap is given by
\begin{equation}
\epsilon^+_j\simeq \frac{\Delta}{k_FR}(j+1/2).
\end{equation}
Analogously, negative energy eigenstates are found for the choice $\kappa=-j-1/2$, $l_e=\ell-1$, and $l_h=\ell$, 
so that we can write $f(r)=Aj_{\ell-1}(kr)$ and $g(r)=Bj_{\ell}(kr)$, with $\ell=1,2,\ldots$ and find the subgap spectrum 
\begin{equation}
\epsilon^-_j\simeq -\frac{\Delta}{k_FR}(j+1/2). 
\end{equation}
For every $j$ we have $2j+1$ degenerate states labelled by $j_z$. The spectrum is shown in Fig.~2a) of the Main text, and it is 
consistent with the generic spectrum of the Dirac equation on the surface of a sphere \cite{ParentePRB2011,GonzalezPRL1992}. 
We see that the spectrum is gapped by an amount $4\Delta/k_FR$, as a result of the finite size of the system and that the gap 
does not depend exponentially on the size of the system, contrary to the case of the planar geometry, but only algebraically. 
It can be shown that the origin of the gapped spectrum is the spin connection that appears in the surface Hamiltonian on a 
curved surface.

\section{Coupling to the magnetic field}
\label{App:CouplingToField}

We now study the full coupling of the Majorana modes localized at the surface of a sphere of radius $R$ to the external field, 
by first separately considering the orbital and the Zeeman term, and then their joint action. 

\subsection{Orbital term}

Neglecting the ${\bf A}^2$ term, and choosing the 
gauge ${\bf A}=\hat\phi A_\phi({\bf r})$, the magnetic field couples via a term
\begin{equation}
{\cal H}_{{\bf p}\cdot{\bf A}}=\frac{e}{m}p_\phi A_\phi({\bf r})
\end{equation}
with $p_\phi=-\frac{i\hbar}{r\sin\theta}\partial_\phi$. The vector potential can be chosen to have the form 
$A_\phi({\bf r})=A(r)\sin\theta$. Due to the spin-triplet pairing the azimuthal momentum $p_\phi$ is not a 
constant of motion. We write the $z$-component of the angular momentum as 
$L_z\equiv -i\hbar\partial_\phi=\hat{J}_z-\hbar s_z/2$, so that we can separate the ${\cal H}_{{\bf p}\cdot{\bf A}}$ 
in a term proportional to $\hat{J}_z$ and a term proportional to $s_z$,
\begin{equation}
{\cal H}_{{\bf p}\cdot{\bf A}}=\frac{e}{m}\frac{A_\phi({\bf r})}{r\sin\theta}\left(\hat{J}_z-\frac{\hbar}{2}s_z\right).
\end{equation}   
For large $j_z$ we can neglect the term proportional to $s_z$. In the limit $R\gg \lambda\gg\xi$ 
we can approximate the matrix elements of the term proportional to $J_z$ as
\begin{equation}\label{Eq:DiracShpereOrbital}
\left\langle\Psi_{j,j_z}^{\alpha}\left| \frac{e}{m}\frac{A(r)}{r}J_z\right|\Psi_{j,j_z}^{\alpha}\right\rangle
=-j_z\frac{\hbar^2}{2m}\frac{2\pi H}{\Phi_0}\frac{\lambda}{R}.
\end{equation}
We thus see that the magnetic field produces a tilting of the spectrum, in analogy with the case of the 
planar and cylindrical surface. The modified spectrum at large $j_z$ is given by
\begin{equation}\label{TiltedDiracSphereLargeJz}
\epsilon^\pm_{j,j_z}= \pm \frac{\Delta}{k_FR}j-j_z\frac{\hbar^2}{2m}\frac{2\pi H}{\Phi_0}\frac{\lambda}{R},
\end{equation}
The energy of the states with $j_z>0$ is lowered, while the energy of the states with $j_z<0$ is raised. 
The threshold field is then given by
\begin{equation}
H^*=\frac{\Phi_0}{2\pi}\frac{2}{\lambda\xi}.
\end{equation}
that is consistent with the results of planar and a cylindrical geometry.

For small values of $j_z$ the contribution proportional to $s_z$ becomes important. The spin operator 
$s_z$ does not commute with the operator $\hat{K}=\tau_z({\bf s}\cdot\hat{\bf L}+\hbar)$, so that it will in 
general couple states at positive energy with states at negative energy. Furthermore, since $s_z$ is even 
under parity it can only couple states differing by an odd number of angular momentum quanta. In particular, 
we find that $s_z$ has finite matrix elements between states of positive energy and quantum numbers 
$(j,j_z)$, and states of negative energy with quantum numbers $(j\pm 1,j_z)$.

Assuming that in the limit $\xi/\lambda\ll 1$ the field does not substantially vary in the region $R-\xi<r<R$, 
where the radial integral takes its largest contribution due to the localization of the Majorana wavefunction, 
it can be taken outside the integral and the relevant matrix elements amount to
\begin{equation}\label{Szcoupling}
\langle\Psi^+_{j,j_z}|s_z|\Psi_{j\pm 1,j_z}^-\rangle=-
\frac{\sqrt{(j+1/2\pm 1/2)^2-j_z^2}}{j+1/2\pm 1/2},
\end{equation}
providing a finite coupling between positive and negative energy states that increases the gap in the spectrum.

\subsection{Zeeman term}
\label{App:Zeeman}

We now consider the action of the Zeeman term. In the spherical geometry its surface contribution reads
\begin{equation}
H_Z=g\mu_B\frac{3H}{2}\left[\sin\theta~\hat{\boldsymbol{\theta}}-\frac{\lambda}{2R}\cos\theta ~\hat{\bf r}\right]\cdot{\bf s}.
\end{equation}
Making use of the identity
\begin{equation}
\hat{\bf r}\cdot{\bf s}\left|{\cal Y}^{j_z}_{j,j\pm 1/2}\right\rangle=-\left|{\cal Y}^{j_z}_{j,j\mp 1/2}\right\rangle,
\end{equation}
together with $\{\hat{\boldsymbol{\theta}}\cdot{\bf s},\hat{\bf r}\cdot{\bf s}\}=0$ we see that the matrix element of the polar 
spin component $\hat{\boldsymbol{\theta}}\cdot{\bf s}$ are approximately zero. More precisely one has
\begin{widetext}
\begin{eqnarray}
\left\langle\Psi^+_{j,j_z}\right|\sin\theta ~\hat{\boldsymbol{\theta}}\cdot{\bf s} \left|\Psi^-_{j'.j'_z}\right\rangle
&=&\delta_{j_z,j'_z}\int_0^Rdr r^2(f_j^+(r)^*f^-_{j'}(r)-g_j^+(r)^*g^-_{j'}(r))
\left[c^{0}_{j,j_z}\delta_{j,j'}+c^{+}_{j,j_z}\delta_{j',j+2}+c^{-}_{j,j_z}\delta_{j',j-2}\right].
\end{eqnarray}
\end{widetext}
In fact, although the angular integral has non-zero matrix elements for $j'=j$ and $j'=j\pm 2$, the radial integral is 
strongly suppressed. It follows that this component has no effect on the Majorana modes, that are found to be 
insensitive to in-plane fields. 

Close to the poles, the field points about the radial direction, that is orthogonal to the surface and fully couples to 
the Majorana modes. Its magnitude is weakened by a factor $\lambda/R$, that makes it comparable to the orbital 
term proportional to $s_z$. Furthermore, the angular matrix elements of the relevant part of the orbital and Zeeman 
terms are the same and an assessment of their relative sign and magnitude becomes crucial.

\subsection{Orbital and Zeeman term}
\label{App:OrbitalZeeman}

The full coupling with the magnetic field is written as
\begin{equation}
H_B=\frac{e}{m}A_\phi({\bf r})p_\phi+g^*\mu_B{\bf B}({\bf r})\cdot{\bf s},
\end{equation}
where $g^*=g/2$ is the ratio between the material $g$-factor and the bare electron's one. Recalling that 
$A_\phi({\bf r})=A(r)\sin\theta$, $p_\phi=-i\hbar\partial_r/(r\sin\theta)$, $-i\partial_r=j_z-s_z/2$, and 
$\mu_B=e\hbar/(2m)$ we have
\begin{eqnarray}
H_B&=&\frac{e\hbar A(r)}{mr}j_z-\mu_B\left(\frac{A(r)}{r}s_z-g^*{\bf B}({\bf r})\cdot{\bf s}\right)\\
&=&\frac{e\hbar A(r)}{mr}j_z-g^*\mu_B\frac{A(r)}{r}\left(1+\frac{A'(r)r}{A(r)}\right)\sin\theta~\hat{\boldsymbol{\theta}}\cdot{\bf s}\nonumber\\
&-&\mu_B\frac{A(r)}{r}\left(s_z-2g^*\cos\theta~\hat{\bf r}\cdot{\bf s}\right).
\end{eqnarray}
In the first line of the second equation we recognize the tilting term and the parallel-to-the-surface polar term. 
The relevant part of the magnetic field that has non vanishing matrix elements between positive and negative 
energy states is given by the third term. We are then left with the calculation of the matrix elements of 
\begin{equation}\label{Eq:M}
M_{j,j'}=-\mu_B\frac{3H\lambda}{2R}\left\langle\Psi^+_{j,j_z}\left|s_z-2g^*\cos\theta~\hat{\bf r}\cdot{\bf s}\right|\Psi^-_{j',j_z}\right\rangle.
\end{equation}
Both the terms $s_z$ and $\cos\theta~\hat{\bf r}\cdot{\bf s}$ are even under parity, so that $j$ and $j'$ has to differ by a 
unit of angular momentum. Defining 
\begin{eqnarray}
u_{j,j'}&=&\int_0^Rdr r^2f_j^+(r)^*f^-_{j'}(r),\\
v_{j,j'}&=&\int_0^Rdr r^2g_j^+(r)^*g^-_{j'}(r)
\end{eqnarray}
where we extended the radial integral to the entire $r<R$ region, we have
\begin{widetext}
\begin{eqnarray}
\left\langle\Psi^+_{j,j_z}\left|s_z\right|\Psi^-_{j'.j_z}\right\rangle
&=&u_{j,j'}\int d\Omega~\left({\cal Y}^{j_z}_{j,j+1/2}\right)^*~s_z~{\cal Y}^{j_z}_{j',j'-1/2}+v_{j,j'}\int d\Omega~\left({\cal Y}^{j_z}_{j,j-1/2}\right)^*~s_z~{\cal Y}^{j_z}_{j',j'+1/2},\\
&=& -u_{j,j'}\frac{\sqrt{(j+1)^2-j_z^2}}{j+1}\delta_{j',j+1}-v_{j,j'}\frac{\sqrt{(j'+1)^2-j_z^2}}{j'+1}\delta_{j'+1,j}\\
\left\langle\Psi^+_{j,j_z}\left|\cos\theta ~\hat{\boldsymbol{r}}\cdot{\bf s} \right|\Psi^-_{j'.j'_z}\right\rangle
&=&-(u_{j,j'}+v_{j,j})\int d\Omega~\cos\theta~\left({\cal Y}^{j_z}_{j,j+1/2}\right)^*~{\cal Y}^{j_z}_{j',j'+1/2}\\
&=&-\frac{u_{j,j'}+v_{j,j'}}{2}\left[\frac{\sqrt{(j+1)^2-j_z^2}}{j+1}\delta_{j',j+1}+\frac{\sqrt{(j'+1)^2-j_z^2}}{j'+1}\delta_{j'+1,j}\right]
\end{eqnarray}
\end{widetext}
Approximating $u_{j,j-1}\simeq v_{j,j+1}\simeq 1/2$, we see that the orbital term is completely 
canceled by part of the Zeeman term and the relevant matrix elements amount to
\begin{equation}
\langle\Psi_{j,j_z}|H_B|\Psi_{j\pm 1,j_z}\rangle=\tilde{g}\mu_B\frac{3H\lambda}{2R}\frac{\sqrt{(j+1/2\pm 1/2)^2-j_z^2}}{j+1/2\pm 1/2}.
\end{equation}
with $\tilde{g}=g^*-1/2$. It follows that the two terms do not cancel completely and that a relevant perturbation results from 
the combined action of the two terms, providing a relevant perturbation to the Majorana modes.

\section{Eigenfunctions of the Surface Dirac operator}
\label{App:DiracSurface}

Here we provide the eigenfunctions of the 2D Dirac operator of the surface of the sphere. For simplicity 
we set velocity to one $v_\Delta=1$. The Dirac operator is given by Eq.~(13) in the main text and we repeat 
it here for convenience. Redefining the Pauli matrices appearing in the main text as 
$(s_x,s_y,s_z)\to (-\sigma_y,\sigma_x,\sigma_z)$ the Dirac operator reads
\begin{equation}
H=-i\sigma_x\left(\partial_\theta+\frac{\cot\theta}{2}\right)-i\frac{\sigma_y}{\sin\theta}\partial_\phi.
\end{equation}
The eigenfunctions are provided in Ref.~\cite{AbrikosovJr} and we refer to that reference for details and 
conventions.  The Dirac operator on the sphere $S^2$ is invariant under transformation of the SU(2) group, 
whose generators are given by
\begin{eqnarray}
\hat{L}_z&=&-i\partial_\phi,\\
\hat{L}_+&=&e^{i\phi}\left(\partial_\theta+i\cot\theta~\partial_\phi+\frac{\sigma_z}{2\sin\theta}\right),\\
\hat{L}_-&=&e^{i\phi}\left(\partial_\theta-i\cot\theta~\partial_\phi-\frac{\sigma_z}{2\sin\theta}\right),
\end{eqnarray}
that satisfy standard commutation rules of the SU(2) algebra, $[\hat{L}_z,\hat{L}_+]=\hat{L}_+$, 
$[\hat{L}_z,\hat{L}_-]=-\hat{L}_-$, $[\hat{L}_+,\hat{L}_-]=2\hat{L}_z$. 
The Casimir operator $\hat{L}^2$ takes eigenvalues $l(l+1)$, with half-integer $l=1/2, 3/2,\ldots$. 
Eigenstates can be labeled by the quantum number $l$ and $m$, with $m=-l,\ldots,l$ eigenvalue 
of $\hat{L}_z$. The eigenfunctions of the Dirac operator are specified by integer 
eigenvalues 
\begin{equation}
\lambda=\pm(l+1/2). 
\end{equation}
We then introduce integers $l_\pm=l\pm 1/2$ and $m_\pm=m\pm 1/2$ and write the eigenfunctions as ($x=\cos\theta$)
\begin{widetext}
\begin{equation}
\Upsilon^\pm_{lm}(x,\phi)=i^l_+(-1)^{\frac{1}{2}(m_-+|m_-|)}\frac{\sqrt{(l+m)!(l-m)!}}{2^{|m|+1/2}\Gamma(l_+)}
\frac{e^{im\phi}}{\sqrt{2\pi}}\left(
\begin{array}{c}
\sqrt{\rho^{(|m_-|,|m_+|)}(x)}P_{l-|m|}^{(|m_-|,|m_+|)}(x)\\
{\rm sign}(m\lambda )\sqrt{\rho^{(|m_+|,|m_-|)}(x)}P_{l-|m|}^{(|m_+|,|m_-|)}(x)\\
\end{array}
\right),
\end{equation}
\end{widetext}
where $\rho^{(\alpha,\beta)}(x)=(1-x)^\alpha(1+x)^\beta$ is the weight function for the Jacobi 
polynomials $P^{(\alpha,\beta)}$ \cite{Gradshteyn}. The eigenstates are normalized as 
$\langle \Upsilon_{lm}^\sigma|\Upsilon_{l'm'}^{\sigma'}\rangle=\delta_{\sigma,\sigma'}\delta_{l,l'}\delta_{m,m'}$. 

\subsection{Matrix elements of the spin connection}

We now have all the elements to calculate the matrix elements of the spin connection
\begin{equation}
\Gamma_\phi=-i\frac{\sigma_z}{2}\cos\theta.
\end{equation}
The matrix elements can be obtained via application of the fundamental identity \cite{Gradshteyn}
\begin{eqnarray}
&&2(n+1)(n+\alpha+\beta+1)(2n+\alpha+\beta)P^{(\alpha,\beta)}_{n+1}(x)\nonumber\\
&=&\left[(2n+\alpha+\beta)(2n+\alpha+\beta+2)x+\alpha^2-\beta^2\right]\nonumber\\
&\times&(2n+\alpha+\beta+1)P_n^{(\alpha,\beta)}(x)\nonumber\\
&-&2(n + \alpha)(n + \beta)(2n + \alpha + \beta + 2) P_{n-1}^{(\alpha,\beta)}(x).
\end{eqnarray}
Together with the matrix elements provided in the Main text it is possible to show that 
\begin{eqnarray}
\langle\Upsilon^\sigma_{lm}|x\sigma_z|\Upsilon^\sigma_{l'm'}\rangle&=&\delta_{mm'}\delta_{l',l}\frac{m}{2l(l+1)},
\end{eqnarray}
that slightly corrects the tilting term.

\bibliography{Bibfile}

\end{document}